\documentclass[pra,aps,amsmath,amssymb,twocolumn,superscriptaddress]{revtex4-1}
\usepackage{color}
\usepackage{graphicx}
\usepackage{ulem}
\newcommand{\bc}{\begin{center}}
\newcommand{\ec}{\end{center}}

\newcommand{\bla}{\color{black}}

\begin{document}
\title{Exploring the role of Leggett-Garg inequality for quantum cryptography}

\author{Akshata       Shenoy      H.}
\email{akshata@ece.iisc.ernet.in}
\affiliation{Elec. Comm. Engg. Dept, IISc, Bangalore, India.}

\author{S.  Aravinda} \email{aru@poornaprajna.org}

\author{R.  Srikanth} \email{srik@poornaprajna.org}
\affiliation{Poornaprajna    Institute    of   Scientific    Research,
  Sadashivnagar, Bangalore, India.}

\author{Dipankar  Home} \email{dhome@jcbose.ac.in} \affiliation{CAPSS,
  Dept.  of Physics,  Bose Institute, Salt Lake Campus,  Kolkata - 700
  091, India.}

\begin{abstract}
In the cryptographic context, an earlier unexplored application of the
temporal version  of the Bell-type  inequality is shown here  in the
device-independent  (DI)   scenario.   This  is  done   by  using  the
Leggett-Garg  inequality (LGI)  to  demonstrate  the security  against
eavesdropping in a  quantum key distribution (QKD)  scheme. This 
typically involves a
 higher dimensional attack against  which the standard
BB84 protocol is insecure. For  this purpose, we invoke an appropriate
form  of  LGI.   While  the  key  generation  is  done  by  the  usual
Bennett-Brassard 1984 (BB84) method, the security check against device
attacks is  provided by  testing for the  violation of  the particular
form of LGI used here. 
\end{abstract}


\maketitle

\textit{Introduction.}  In quantum key distribution (QKD), two distant
parties (Alice  and Bob) securely  share a private random  bit string,
whose  security  against  an  eavesdropper Eve  is  based  on  quantum
features   like  no-cloning   and   imperfect  distinguishability   of
non-orthogonal states.  A  QKD protocol was first  proposed by Bennett
and Brassard  (BB84) \cite{BB84}.  In  a subsequent work,  Ekert (E91)
\cite{Eke91} showed how quantum  entanglement and quantum violation of
Bell-type inequality (BI) \cite{Bel64,CHS+69} could be used as a basis
for QKD.  The intuition behind achieving security in this way was that
Eve's intervention  would tend to disentangle  the particles.  Spurred
on  by the  E91 protocol,  Bennett, Brassard  and Mermin  \cite{BBM92}
proposed a  different entanglement-based  scheme that did  not require
Bell's theorem and showed its  equivalence with BB84. In the following
years, more sophisticated security proofs were provided in the context
of a  variety of attacks  on the transmission channel,  culminating in
the proof of unconditional security of BB84 \cite{SP00}.

However,  in  recent  years,  it  has been  realized  that  the  above
mentioned security proofs of QKD schemes have limited practical value,
because of  their implicit  assumption that all  the devices  used for
state   preparation    and   measurement   are    well   characterized
\cite{esther}. For  example, in  BB84, the malicious  eavesdropper Eve
may herself  be the vendor, who  supplies states and devices  to Alice
and  Bob. She  may  give  them higher  dimensional  states as  ``cheat
states'' instead of those legitimate  for the protocol, and access the
extra dimensions,  as shown below,  to compromise the security  of the
protocol.   This type  of attack  is  known as  ``device attack'',  as
opposed  to the  usual  ``channel attack''  where  Eve intercepts  the
transmitted states.  Thus, a QKD  protocol must be secure against both
channel  and device  attacks.   While BB84  is unconditionally  secure
under    channel    attacks,    it    is    not    secure    in    the
\textit{device-independent} (DI)  scenario, i.e.  one in  which device
attacks are allowed.
 
It is in this  context that we point out how  the temporal analogue of
the Bell-type  inequality, known as the  Leggett-Garg inequality (LGI)
\cite{LG85, Leg02}, can be relevant  in formulating QKD schemes in the
DI scenario.   The assumptions leading  to LGI  are that a  system has
definite  values of  the observable  properties at  \textit{any} given
instant,  and  that  these  values  can  be  accessed  by  noninvasive
measurements. LGI imposes an upperbound on a linear combination of the
correlations  $P(a,b|x,y)$ (where  $a  (b)$ are  outcomes obtained  on
measuring observables $x (y)$) between the outcomes of measurements of
the observables at different instants  that are obtained by sequential
measurements  on  the  \textit{same}  particle.   While  the  original
motivation leading  to LGI was for  testing the validity of  QM in the
macro-limit, in recent years, LGI has been applied in various contexts
using different  micro-systems, resulting  in a number  of theoretical
\cite{KB07} and experimental
\cite{WHW+00,*RKM06,*FAB+11,*ARM11}  works.   In  the  light  of  this
upsurge of interest  about LGI, it should be of  significance to probe
whether LGI can have an application for ensuring security in a quantum
cryptographic  task.    This  specific   question  has   yet  remained
unexplored.  We note  that the only other prior application  of LGI in
the area  of quantum information  was for  saving memory in  a quantum
information processing task \cite{BTC+04}.  Before proceeding to apply
LGI in  the context considered  in our paper,  we note that  the usual
form of  LGI involves temporal  correlations present in the  two state
oscillations. For the purpose of our paper we consider the form of LGI
proposed by Brukner et al. \cite{BTC+04}.

The plan  of the paper  is as follows. We  begin by discussing  the DI
scenario involving a  typical higher dimensional attack, which
we call the  AGM attack \cite{AGM06}.  Next, the suitable  form of LGI
which can  be of cryptographic  use is introduced.  We then discuss  
the limitation of a LG-protocol  which is the direct temporal analogue
of the  CHSH protocol \cite{AGM06}.   Subsequently,  we propose a
\textit{LG-BB84}  protocol  to  circumvent the  limitation  affecting  the
LG-protocol. In  the LG-BB84 protocol,  the key generation is  done by
the BB84  method using mutually  unbiased basis for bit  encoding. The
security of  BB84 against general  channel attacks has been  proved in
\cite{M01,SP00}. But, BB84 is not secure against device attacks,
  in particular the AGM attack. We address the insecurity against this
  particular widely discussed device attack  and detect the AGM  attack by introducing
an  additional  basis  for  Bob which  establishes  the  required  LGI
violating correlations  between Alice and  Bob. Thus, the  security of
LG-BB84 in this DI scenario  is proven by  deriving a
positive secret key rate under a combination of individual channel  attack and
 AGM device  attack.  Even at lower error rate,  compared to the
original scheme,  a threat from  the AGM attack may  exist, which is  detected by
a reduction in the LGI violation.  The  paper is concluded by   
summarizing the salient features of our work and indicating the direction
for further work.

\textit{Device-independence.}  In BB84 Alice transmits
states randomly prepared in the conjugate basis (Eg: Pauli spin basis,
$X$ or $Z$)  and Bob's random measurements in one  of these two basis.
Over a classical  channel, they determine the cases  where their bases
match, discarding  the rest.  On  a smaller, randomly  selected sample
from  the retained  cases,  Alice and  Bob  announce their  respective
outcomes to compute the error  rate.  Unless this rate is sufficiently
small, they abort the protocol run.  The statistics that arise in BB84
are: $ P(a=b|x=y) = 1; P(a=b|x\ne y) = \frac{1}{2}.$ If Eve intercepts
Alice's   transmission  to   acquire  information,   because   of  the
information-vs-disturbance trade-off, she inevitably disrupts the BB84
statistics, which is detected by  Alice and Bob.  This constitutes the
essential security of BB84.

Here it  is implicitly  assumed that  Alice and  Bob know  exactly how
their  correlations $P(a,b|x,y)$  are produced;  in other  words, that
their  devices  are  trustworthy,  and that  they  are  measuring  the
properties of the \textit{same} particle.  In the DI scenario, Eve can
cheat by having  them measure different particles  using, for example,
the  following  ploy.  Eve  presents  them  with the  separable  state
\cite{AGM06}
\begin{equation}
 \rho_{\mathcal AB} = \frac{1}{4}\left(\Pi_{00}^{(12)} +
 \Pi_{11}^{(12)}\right)\otimes\left(\Pi_{++}^{(34)} +
 \Pi_{--}^{(34)}\right)
\label{eq:cheat}
\end{equation}
where $\Pi_{xy}$ indicates projector  to the state $|x,y\rangle$.  The
bracketed superscripts  in the definition of  $\rho_{\mathcal AB}$ are
particle labels.  Eve has so  arranged the devices such that particles
1 and 3  (2 and 4) are  with Alice (Bob).  When Alice  and Bob measure
$Z$ ($X$), they measure particles 1 and  2 (3 and 4). Notice that this
reproduces the BB84  statistics, but after the  public announcement of
basis by  Alice and Bob, Eve  has the `hidden variable'  pertaining to
which pair of particles Alice and  Bob measure, she can find out their
secret bit (0 or 1)  with certainty without introducing a disturbance.
Eve's cheating  here hinges  on the  fact that  Alice and  Bob believe
their system to  be a single system of dimension  two (a qubit), while
in fact they  are accessing a system of higher  (= 16) dimensionality.
Thus, it becomes crucial to rule  out hidden dimensions of the Hilbert
space describing the quantum systems  used for QKD. 

The necessary condition for  security in this device-independent (DI)
scenario is that the correlations $P(a,b|x,y)$ shared between Alice and
Bob should satisfy the following inequality,
\begin{equation}
P(a,b|x,y) \ne \sum_\lambda P(a|x,\lambda)P(b|y,\lambda)p_\lambda,
\label{eq:factorizable}
\end{equation}
since otherwise it is possible that Eve possesses a copy of the hidden
variable  $\lambda$, and  determines Alice's  and Bob's  outcomes when
they publicly  announce $x$ and  $y$.  This implies  that $P(a,b|x,y)$
must violate a  correlation inequality like BI for  security and hence
Alice and  Bob must share quantum  entanglement.  It turns out  that a
sufficiently large violation  of BI for correlations  that are between
spatially separated  particles guarantees security not  just against a
quantum mechanical Eve,  but even by a general Eve  restricted only by
the  no-signaling condition  (Alice's  measurement  settings or  input
should not influence Bob's  output)\cite{BHK05,AGM06}. A crucial point
to note here is that the BB84 statistics do not violate BI and thus is
insecure in the DI scenario.

\begin{figure}
\includegraphics[width=8.7cm]{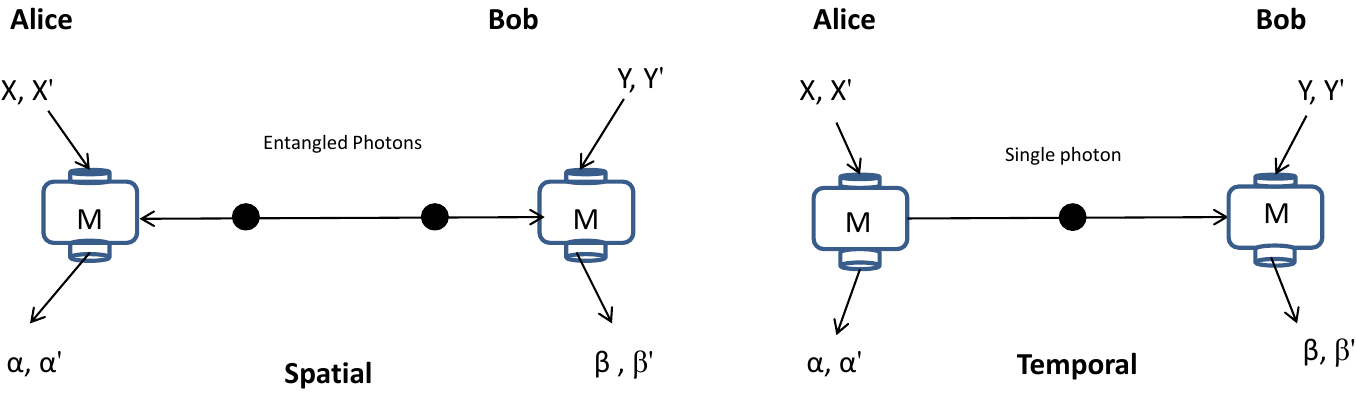}
\caption{Temporal  vs  spatial   correlations  considered  in  the  DI
  scenario.  In the temporal  case, Alice  and Bob  perform sequential
  measurements  on  the same  particle  using different  non-commuting
  observables  at different  times  to obtain  LGI  violation. In  the
  spatial case,  Alice and  Bob measure spatially  separated entangled
  particles using commuting observables at the same instant.}
\label{fig:LG}
\end{figure}

\textit{DI scenario using LGI.} The form of LGI used in our discussion
is given by \cite{BTC+04},
\begin{equation}
\Lambda \equiv 
|\langle x_{t_1}y_{t_2} + x_{t_1}y'_{t_2} + x'_{t_1} y_{t_2} -x'_{t_1}y'_{t_2}
\rangle|\le 2
\label{eq:chsh}
\end{equation}
where at an instant $t_1$ $(t_2)$ where $t_2>t_1$, Alice (Bob) may choose to measure
the  observable  $x$ $(y)$  or  $x^\prime$  $(y^\prime)$.  Unlike  the
original LGI which  involves measuring the same  observable in the presence of time
evolution of  the system  under consideration, the  form of  LGI given
here involves  measuring different observables without  time evolution
of the system. Like the original form of LGI \cite{LG85}, the
inequality given by Eq.(\ref{eq:chsh}) is also based on the same assumptions of realism
and noninvasive  measurability \cite{KB08} and can be considered as the temporal
version of BI which is violated by quantum mechanics. In what follows, we will show
how the incorporation of a LGI test based on  testing the violation of the above inequality can secure
the BB84 protocol against the AGM attack even without the use of entanglement, where the observed reduction in the 
violation of the LGI occurs depending on the amount of device tampering by Eve.
Here we note that within the framework of BB84 scheme, in order to make use 
of the causal ordering of the events of Alice and Bob for ensuring security in the DI
scenario, it becomes nescessary and natural to invoke the violation of LGI-type inequality
that results in the required security contingent on the assumption of no unauthorized signal leakage.
In the present paper since it focussess on initiating a new direction of study in QKD,
it suffices to consider a typical device attack such as the AGM attack \cite{AGM06}, and in this context
we not consider more general device attacks \cite{BCK12,PawB11} which will be studied in a sequel paper.
Before proceeding to discuss our prpoposed protocol, it may be useful to refer to
Fig.   \ref{fig:LG} which illustrates  the  difference  between the settings involving spatial  and
temporal correlations considered in the DI scenario. \bla

If  one wishes  to  have  an  \textit{LG  protocol}, the  temporal
analogue of the CHSH protocol, then Alice and Bob have the  
choice of measuring two dichotomic observables
$x,x^\prime$ and  $y,y^\prime$ respectively, whose  outcome statistics
can be used to test LGI violation.  After many trials, Alice announces
her  settings ($x$  or $x^\prime$)  and Bob  keeps his  outcomes as-is
for all cases except for the  last setting ($x^\prime, y^\prime$), in  which case he
flips his outcome.  This step  is known as  basis reconciliation.
The violation of  LGI (Eq. (\ref{eq:chsh})) would ensure that with high
probability  their  data  has  positive  correlation  (i.e.,  is  more
correlated  than anti-correlated).  To share an  identical secure key  they
proceed  with further  classical cryptographic post-processing.   As the  cheat state
$\rho_{AB}$  cannot  violate  LGI,  this  protocol  offers  a  natural
protection against the higher  dimension attack.  

Note that an LG protocol obtained as the temporal analogue of the CHSH
protocol \cite{AGM06} may seem to have the same security implications.
However, we will find that the temporal correlations are characterized
by invasiveness  (or ``signaling'', in the  framework of correlations)
which results  in a weakening of  the monogamy bound. What  we mean by
this  is   that,  given  the  sequential   measurements  for  temporal
correlations by Alice, followed by Eve and then Bob, we find (Appendix
A)
\begin{equation}
\Lambda_{\mathcal  AE}  +   \Lambda_{\mathcal  AB}  \leq  2\sqrt{2}  +
\sqrt{2} = 3\sqrt{2}.
\label{eq:mon}
\end{equation}
which is larger than 4, the no-signaling bound applicable to the spatial correlations \cite{Ton06}.
The above relation implies that if Alice-Eve correlation violates LGI, then a subsequent measurement by Bob
cannot establish the same amount of correlation between Alice-Bob. Eve's intervening measurement during the particle's
channel transit renders Alice-Bob correlations separable (where $\sqrt{2}$ is the separable bound).
Note that if   the  measurements   are   `anchored'  on   $E$,   then  we   have
\begin{equation}
\Lambda_{\mathcal AE} + \Lambda_{\mathcal BE} \leq 2 \times 2\sqrt{2}
=  4\sqrt{2}
\end{equation} 
as already observed in \cite{BTC+04}. This implies that Eve can simultaneously
share the LGI violating correlations between Alice and Bob. This demonstrates the weakening  of monogamy
in the case of temporal correlations and has the  cryptographic
implication  that,  to  be  assured  of  security,  a  larger  degree
violation of LGI would be required  in the LG protocol than the degree
of violation  of the  CHSH inequality required  in the  CHSH protocol. 
However, the LG protocol is conceptually important 
because it  highlights the  fundamental  difference between
spatial and temporal correlations, from the perspective of security. 

%\begin{widetext}
%\ec
%\end{widetext}

\textit{LG-BB84  protocol.}  The  discussion in  the previous  section
suggests that it would be advantageous to decouple the device attack detection 
(LGI test affected by the weakening of monogamy)
from the key generation (which is done by the BB84 scheme) and channel attack detection
(which is independent of LGI violation and hence of weakening of monogamy). For this purpose, to establish the LGI
violating correlations between Alice-Bob, we introduce an additional basis $M_\pm  \equiv \frac{1}{\sqrt{2}}(X  \pm Y)$
at Bob's end. Based on these considerations, we propose the LG-BB84 protocol as follows:(1) Alice
prepares states  randomly in $X$  or $Y$  basis; (2) Bob  measures the
incoming photons  randomly in $X$,  $Y$ or $M_\pm$ basis;  (3) After
Bob classically acknowledges  receipt, Alice  announces her preparation  bases, as
does Bob  his measurement bases.  (4) In  cases where their  bases are
matched,  the resulting  outcomes constitute  the raw  key.  If  their
bases are  mismatched and Bob measured  in the $X$ or  $Y$ basis, the
outcomes are discarded. If their bases are mismatched and Bob
measures $M_\pm$,the outcomes are used to test for violation of the following
form of LGI obtained by rewriting Eq.(\ref{eq:chsh}) where $x(x')$ and $y(y')$ are
replaced by $X(Y)$ and $M_+(M_-)$ respectively:
\begin{equation}
\Lambda_{AB}      \equiv      |\langle     X_{t_1}{M_{+}}_{t_2}      +
X_{t_1}{M_{-}}_{t_2}  +  Y_{t_1}{M_{+}}_{t_2}  -  Y_{t_1}{M_{-}}_{t_2}
\rangle|\le 2,
\label{eq:lgm}
\end{equation}
where Alice has the choice of preparing the photon state in $X$ or $Y$
at time $t_1$ and Bob measures the corresponding state in either $M_+$
or $M_-$ basis at time $t_2$ ($t_2  > t_1$).  Eve intervenes at a time
  $t_1^\prime$ ($t_1<t^\prime_1<t_2)$.   The cheat  state cannot
pass  the LGI  test  since ($\Lambda_{AB}(\rho_{AB})=0$),  essentially
because LGI-violating  temporal correlations cannot be  established by
measuring two  different particles. The  LGI test thus serves  like an
identity  check  on   Bob's  particle,  certifying  that   it  is  the
\textit{same} particle that Alice transmitted.

Let us consider an
eavesdropping model   which  is  a   combination  of  device- and  channel-attacks. 
For  the device attack,  Eve mixes a  fraction $f$  of the
cheat states $\rho_{\mathcal AB}$ with the legitimate BB84 states.  For the channel attack,
Eve intercepts individual qubits and makes them interact with her  ancillary qubit via the following unitary
evolution \cite{FGG+97}:
\begin{eqnarray}
U|00\rangle_{BE} &=& |00\rangle_{BE} \nonumber \\
U|10\rangle_{BE} &=& \cos\theta|10\rangle_{BE} + \sin\theta|01\rangle_{BE}.
\label{eq:NG}
\end{eqnarray}
where $\theta  \in [0,\pi/2]$  is the attack  parameter (a  measure of
interaction  between  transmitted qubit-ancilla system).   Eve  then  waits for the public
announcement of bases and measures her ancilla  in the appropriate 
basis corresponding to each transmitted qubit. For 
individual attacks the interaction in Eq.(\ref{eq:NG}) is optimal\cite{BLM96} in the sense that for a
given degree of noise, Eve's mutual  information with Alice $I(A:E)$ is maximal.

Suppose,  $f=0$,  i.e.  the  cheat  states  $\rho_{\mathcal{AB}}$  are
absent.  Then, error rates as seen by Alice-Bob, Alice-Eve and Bob-Eve
due to Eve's channel attack  alone (Appendix B) are respectively given
by,   $e_{AB}=    (1-\cos\theta)/2$,   $e_{AE}=(1-\sin\theta)/2$   and
$e_{BE}=(1-\sin2\theta)/2$, irrespective  of the basis  or transmitted
bit (cf.  \cite{SG01,*SG201}). The condition for  extracting a shorter
secure  key  from  a  longer  raw  key  (using  one-way  communication
protocols for key reconciliation  and privacy amplification), i.e. for
deriving  a positive  key  rate \cite{CK78}  i.e.  $I(A:B)>I_E  \equiv
\min[I(A:E),I(B:E)]$. This criterion in our protocol is satisfied when
$\theta  \le \frac{\pi}{4}$  for  which $e_{AB}  \lesssim 14.6\%$  and
$\Lambda_{AB}  \geq  2\sqrt{2}\cos(\theta)$.   This implies  that  the
protocol  is  secure when  the  error  rate  is  less than  the  above
tolerable  limit and  precisely when  it  is LGI  violating. Thus  the
condition for security is equivalent to the requirement that Alice-Bob
correlation data violates LGI.

\begin{figure}
\includegraphics[width=8cm]{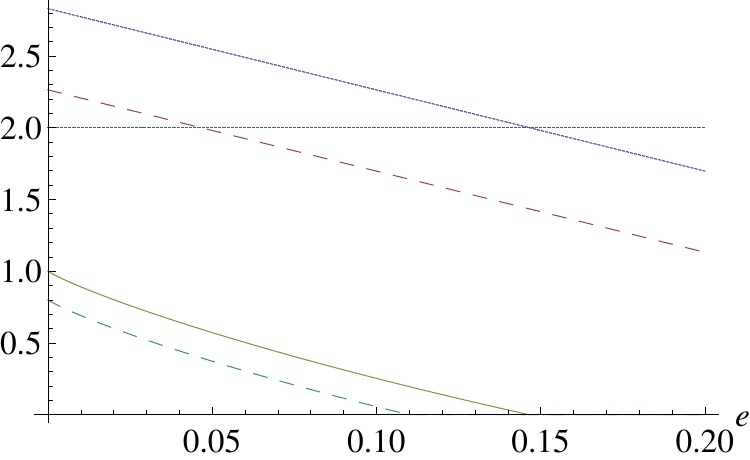}
\caption{The upper  pair of lines represent an LGI  test, while the
  lower pair represent positive key  rate $K \equiv I_{AB}-I_E$, as a
  function of $e  \equiv e_{AB}$.  The solid  (dashed) lines represent
  the  case $f=0$  ($f=0.2$),  where  $f$ is  the  fraction of  higher
  dimension attack by Eve. As $f$ increases, Eve produces less channel
  error for  the same  level of knowledge  about the  transmitted secret
  bits. Device tampering by Eve leads to a reduction in the violation
 of LGI. This, along with the channel attacks, reduces the error rate.} \bla
\label{fig:lg_rate}
\end{figure}

Let $f>0$, corresponding to Eve launching both the channel and device attacks. The
error rate and the LGI violation observed by Alice and Bob gets modified to:
\begin{eqnarray}
e^\prime_{AB} &=& (1-f)e_{AB} = \frac{1}{2}(1-f)(1-\cos\theta).\nonumber\\
\Lambda_{AB} &=& 2\sqrt{2}\cos\theta(1-f).
\label{eq:eAB}
\end{eqnarray}
Correspondingly, Alice-Bob mutual information is given by $I(A:B)=1 -
H\left(e_{AB}^\prime\right)$, while Alice-Eve mutual information is as follows 
\begin{equation}
I(A:E) = (1-f)(1 - H(e_{AE})) + f,
\label{eq:IAE}
\end{equation}
(Eve  knows the  bit value  determinsitically in  fraction $f$ corresponding to the device attack ),
where $H$ is Shannon binary entropy; and similarly for $I(B:E)$.  From
the  above equation,  we observe  that for constant $\theta$, with $f$
increasing, the  protocol becomes insecure  at lower error  rates (see
Figure \ref{fig:lg_rate}). This is because Eve produces fixed channel error due to her
channel attack while  gaining  more information  on  the  secret bits using the
cheat states. From Figure \ref{fig:lg_rate},  we  see that  the  range  $10.9\% \lesssim  e_{AB}
\lesssim 14.6\%$,  which is secure  under the channel attack alone,  
is insecure in the DI scenario considered. Such eavesdropping 
is detected using the LGI violation
and would not be detected otherwise.  Thus,  Alice and
Bob can  determine the highest tolerable  error rate and bound
their key rate only with  the combined BB84 and LGI data, and
solving  for   $\theta$  and  $f$  using   Eqs.   (\ref{eq:eAB}).   

\textit{Summary    and   Concluding    remarks.}   In the non-DI
scenario,  standard BB84 is known to be
secure. In the  DI scenario, the possibility of  mistrusted devices is
allowed. In this case, for the security,  entanglement is believed to be
essential.   As  a result,  all the proposed  QKD protocols  to-date in the DI scenario are
entanglement-based.  In contrast, the  present  paper  shows  the  possibility  of
avoiding the  use of  entanglement for the type of DI scenario  considered in the context of
AGM attack \cite{AGM06}. This is done by augmenting BB84  with a  security check  through the
testing of  the violation  of an inequality (namely,  an appropriate
form  of  LGI)  involving  the  correlations  between  the  results  of
temporally  separated measurements  of suitable  observables. Our  key
demonstration is that,  for the AGM attack by
Eve in the DI scenario,  our protocol  is secure, whereas BB84 is not.

A few points need to be mentioned here.   First, the temporal  correlations
present in LGI are less  monogamous than their spatial counterparts in
BI in  that the right hand side  of Eq. (\ref{eq:mon})  can exceed 4.   As we have
explained earlier, this  makes the LG protocol  less secure than the  CHSH protocol,
highlighting  a   basic  difference   between the spatial   and  temporal
correlations. Second, in the LG-BB84 protocol proposed here, the tolerable error rate depends on the
combined  LGI and  the  BB84  data, while the scheme in itself  may be
considered as  the prepare-and-measure  version of E91  protocol.

Note that in the DI scenario that has been considered, it is assumed that there are
no  state  transmissions   of  cheat  states  and   that  devices  are
memoryless.  If the  former assumption is given up,  then the security can
be  proven in the semi-device-independent scenario
\cite{PawB11}, where the dimensionality of the transmitted state is
bounded      from      above       using      dimension      witnesses
\cite{GBH+10, HGM+12}. If  the latter  assumption is not made,
then memory attacks  can occur  in which devices procured from adversarial suppliers  may reveal information about inputs
and outcomes of earlier runs  through the public communication channel
used in the subsequent runs \cite{BCK12}. Against such general attacks, DI security can be proven\cite{MAS06,QFL+07} provided errors are less than 2\%\cite{VV12}. On the other hand, since our work initiates an entirely new direction of study by exploiting LGI, in order to illustrate its efficacy, we have used, to begin with, a specific typical and widely discussed device attack (AGM Attack). Subsequently, this paper will be followed by sequel works pursuing the extension of this type of novel scheme towards making it secure against more general attacks in the DI scenario.

\acknowledgments

We  thank C.  Brukner, N.   Brunner, V.  Ranjith and  P. Mandayam  for
helpful  comments.   SA  acknowledges   support  through  the  INSPIRE
fellowship [IF120025] by DST, Govt.   of India, and Manipal University
graduate  program.  RS  and DH  acknowledge support  from the  DST for
projects  SR/S2/LOP-02/2012  and SR/S2/LOP-08/2013,  respectively.  DH
thanks Center for Science, Kolkata, for support. 

\bibliography{axta}

%\newpage

\appendix

\section{Monogamy in the case of temporal correlations
\label{sec:mon}} 

Let Alice and Bob establish correlations by sequentially measuring 
observables $\hat{x}$ and  $\hat{y}$ on a given qubit at respective instants
$t_1$ and $t_2$.  The probability for Bob
to obtain  outcome $\beta$  on input  $\hat{x}$, after  Alice obtained
outcome $\alpha$ on input $\hat{x}$ is:
\begin{widetext}
\begin{equation}
P_{\alpha\beta|\hat{x}\hat{y}}  =   \textrm{Tr}\left(\frac{1 +
  \beta\hat{y}}{2} \frac{1 + \alpha\hat{x}}{2}\rho \frac{1 +
  \alpha\hat{x}}{2}\right) = \frac{1}{4} +
\frac{\alpha}{4}\textrm{Tr}(\hat{x}\rho) +
\frac{\beta}{8}\textrm{Tr}(\hat{y}\rho) +
\frac{\beta}{8}\textrm{Tr}(\hat{x}\hat{y}\hat{x}\rho)  +
\frac{\alpha\beta}{8}\textrm{Tr}(\{\hat{x},\hat{y}\}\rho),
\label{eq:qcor}
\end{equation}
\end{widetext}
where  outcomes $\alpha,  \beta  = \pm1$.   Therefore, Bob's  marginal
$P_{\beta|xy}  =   \sum_\alpha  P_{\alpha\beta|xy}  =   \frac{1}{4}  +
\frac{\beta}{8}\textrm{Tr}(\hat{y}\rho)                              +
\frac{\beta}{8}\textrm{Tr}(\hat{x}\hat{y}\hat{x}\rho)$    depends   on
Alice's setting (while the converse  is not true).  This dependence of
Bob's  outcome probability  on  Alice's settings,  which  is simply  a
manifestation of  invasiveness, makes temporal  correlations signaling
[24].   If   the  correlations   $P_{\alpha\beta|\hat{x}\hat{y}}$  are
non-signaling, then the following bound holds [18]:
\begin{equation}
\Lambda_{\mathcal AE}^{\rm Bell} + \Lambda_{\mathcal AB}^{\rm Bell} \le 4.
\label{eq:nosig}
\end{equation}
where  the superscript  ``Bell'' indicates  that the  correlations are
spatial in this case.  Temporal correlations violate this bound.

Now consider the case of consecutive measurements $\hat{x}$, $\hat{e}$
and $\hat{y}$ performed at $t_1$, $t_2$ and $t_3$ ($t_1  < t_2 <
t_3$) respectively:
\begin{eqnarray}
\langle \hat{x}, \hat{y} \rangle &=& \hspace{-0.3cm} \sum_{m,n,o =\pm 1}
\hspace{-0.2cm}   mo~\mbox{Tr}\left[\rho \Pi^{m}_{\textbf{x}}\right]
\mbox{Tr}\left[\Pi^{m}_{\textbf{x}} \Pi^{n}_{\textbf{e}}\right]
\mbox{Tr}\left[\Pi^{n}_{\textbf{e}} \Pi^{o}_{\textbf{y}}\right]
\nonumber  \\  &=&  (\textbf{x}  \cdot \textbf{e})  (\textbf{e}  \cdot
\textbf{y}), \label{eq:dis}
\end{eqnarray}
where $\Pi^m_{\textbf{n}}$ is the projector in general along direction
\textbf{n}  with outcome  $m$.  Eq.   (\ref{eq:dis}) implies  that the
third correlation  is ``temporally'' \textit{disentangled}  from first
[12], when the second measurement  is projective. Thus, if Eve attacks
and  measures  at a  time  $t^\prime_1$  in  between Alice  and  Bob's
measurements,  she   completely  disentangles  them.    The  Alice-Bob
correlation  does not  violate  LGI since  their  states are  rendered
``temporally'' separable.

We note that  the l.h.s of (\ref{eq:dis}) has the  form of measurement
on  the  product state  of  the  identical  copies with  Bloch  vector
\textbf{e}, along directions \textbf{x} and \textbf{e}.  For separable
states  in  quantum mechanics,  it  can  be  shown that  $\Lambda  \le
\sqrt{2}$,  whereas the  local bound  is  2.  The  separable bound  is
saturated  with settings  \textbf{x}$^\prime$, \textbf{y},  \textbf{x}
and \textbf{y}$^\prime$  being coplanar,  separated by  angle $\pi/4$,
with $\textbf{e}  = \textbf{y}$.  As a  result, we have for  the above
sequential  measurements   (Eq.   (4)):  $  \Lambda_{\mathcal   AE}  +
\Lambda_{\mathcal AB} \leq 2\sqrt{2} + \sqrt{2} = 3\sqrt{2}.  $

\section{Error rate and LGI violation in LG-BB84 \label{sec:LG-BB84}}

Suppose    Alice    transmits    the    state    $|{\pm}\rangle_X    =
\frac{1}{\sqrt{2}}(|0\rangle \pm |1\rangle)$.  Subjected to the attack
of Eq. (7), this becomes the state
\begin{equation}
\tau^{\pm}_X(\theta) := \frac{1}{2}\left(\begin{array}{cc}
 1 + \sin^2(\theta) & \pm\cos(\theta) \\
\pm\cos(\theta) & \cos^2(\theta),
\end{array}
\right)
\end{equation}
while Eve's probe is left in the state:
\begin{equation}
\tau^{\prime\pm}_X(\theta)                                    :=
\frac{1}{2}\left(\begin{array}{cc}     1     +    \cos^2(\theta)     &
  \pm\sin(\theta) \\ \pm\sin(\theta) & \sin^2(\theta)
\end{array}
\right).
\end{equation}
Similar  expressions  are obtained  for  Alice's  and  Eve's state  when
 a   $Y$-basis   state is transmitted,    given   respectively   by
$\tau^{\pm}_Y(\theta)$    and    $\tau^{\prime\pm}_Y(\theta)$.  It turns
out that the
probability that Eve's attack leads  to an error is the same
for any of Alice's four preparations, and is given by:
\begin{equation}
e_{AB}        =       {_X\langle}\mp|\tau^{\pm}_X|\mp\rangle_X       =
{_Y\langle}\mp|\tau^{\pm}_Y|\mp\rangle_Y = \sin^2(\theta/2).
\label{xeq:eAB}
\end{equation}

It  can  be  shown   that  Eve's  optimal  projective  measurement  to
distinguish   $\tau^{\prime+}_X(\theta)$  ($\tau^{\prime+}_Y(\theta)$)
from    $\tau^{\prime-}_X(\theta)$   ($\tau^{\prime-}_Y(\theta)$) is  by
measuring her probe in the $X$  ($Y$) basis. To see this, suppose that
Alice announces the transmitted state to  be in the $X$ basis, and let
Eve measure her probe in the  basis having one of the projectors to be
$|\xi\rangle  = \alpha|0\rangle  + \beta  e^{i\gamma}|1\rangle$, where
$\alpha,   \beta$   and  $\gamma$   are   real   numbers.  One   finds
$$\langle\xi|\left(\tau^{\prime+}_X(\theta) - \tau^{\prime-}_X(\theta)
\right)    |\xi\rangle    =    2\alpha\beta\sin(\theta)\cos(\gamma).$$
Maximizing  this quantity, it  follows that  we should  set $\gamma=0$
while $\alpha  = \beta = \frac{1}{\sqrt{2}}$. If  Alice announces $X$,
then  Eve measures  her  probe in  the  same basis,  and likewise  for
Alice's  announcement  of  $Y$.  The  error rate  of  Eve  on  Alice's
transmission is:
\begin{equation}
e_{AE}      =     {_X\langle}\mp|\rho^{\prime\pm}_X|\mp\rangle_X     =
{_Y\langle}\mp|\rho^{\prime\pm}_Y|\mp\rangle_Y                        =
\frac{1}{2}(1-\sin(\theta)).
\label{xeq:eAE}
\end{equation}
The error rate on transmissions between Bob and Eve is obtained from
Eqs. (\ref{xeq:eAB}) and (\ref{xeq:eAE}) and found to be:
\begin{eqnarray}
e_{BE}  &=& e_{AB}\left(1-e_{AE}\right)  + \left(1-e_{AB}\right)e_{AE}
\nonumber \\ &=& \frac{1}{2}(1-\sin(2\theta)).
\label{eq:eBE}
\end{eqnarray}

The LGI  inequality (3) is now  represented by one in  which the first
correlatum  is given  by the  BB84 state  preparation weighted  by its
preparation  probability.   For  example,  the  first  term  in  Ineq.
(3) is given by $P(y=+1|x=+1)P(x=+1) + P(y=-1|x=-1)P(x=-1)
+      P(y=+1|x=-1)P(x=-1)      +      P(y=-1|x=+1)P(x=+1)$.      Thus
Ineq. (3) now becomes:
\begin{widetext}
\begin{eqnarray}
\Lambda_{AB} &=& 
\frac{1}{2}\left(\left[P\left(|{+}\rangle_{M_+}||{+}\rangle_X\right) +
P\left(|{-}\rangle_{M_+}||{-}\rangle_X\right) -
P\left(|{+}\rangle_{M_+}||{-}\rangle_X\right) -
P\left(|{-}\rangle_{M_+}||{+}\rangle_X\right)\right] \right. \nonumber \\
&&+ \left[P\left(|{+}\rangle_{M_-}||{+}\rangle_X\right) +
P\left(|{-}\rangle_{M_-}||{-}\rangle_X\right) -
P\left(|{+}\rangle_{M_-}||{-}\rangle_X\right) -
P\left(|{-}\rangle_{M_-}||{+}\rangle_X\right)\right] \nonumber \\
&&-\left[P\left(|{+}\rangle_{M_+}||{+}\rangle_Y\right) +
P\left(|{-}\rangle_{M_+}||{-}\rangle_Y\right) -
P\left(|{+}\rangle_{M_+}||{-}\rangle_Y\right) -
P\left(|{-}\rangle_{M_+}||{+}\rangle_Y\right)\right]  \nonumber \\
&&+ \left. \left[P\left(|{+}\rangle_{M_-}||{+}\rangle_Y\right) +
P\left(|{-}\rangle_{M_-}||{-}\rangle_Y\right) -
P\left(|{+}\rangle_{M_-}||{-}\rangle_Y\right) -
P\left(|{-}\rangle_{M_-}||{+}\rangle_Y\right)\right]\right) \nonumber \\
&=& 2\sqrt{2}\cos(\theta). 
\label{eq:BAB}
\end{eqnarray}
\end{widetext}
where $|\pm\rangle_{M_\pm}$  are the  eigenstates of $M_{\pm}$  and we
assume  that  $P(|\pm\rangle_X)  =  P(|\pm\rangle_Y)  =  \frac{1}{2}$.
Interestingly,  the quantity  $2\sqrt{2}$  saturates  the upper  bound
obtained   by    the   Horodeckis   criterion   in    the   equivalent
entanglement-based   protocol  [21].    The   quantity  analogous   to
(\ref{eq:BAB})  may   be  calculated   similarly  for   the  Alice-Eve
correlations, which we find to be:
\begin{equation}
\Lambda_{AE} = 2\sqrt{2}\sin(\theta).
\label{eq:BAE}
\end{equation}

\end{document}